\documentclass[sort&compress,final]{aipproc}\layoutstyle{8x11double}
\begin{document}\title{Analytic Quest for Confining Interaction
Kernels in Instantaneous Bethe--Salpeter Equations}
\author{Wolfgang
LUCHA}{address={Institute for High Energy Physics, Austrian
Academy of Sciences, Nikolsdorfergasse 18, A-1050 Vienna,
Austria}}\maketitle

\noindent The Salpeter equation is a well-known three-dimensional
approximation to the Bethe--Salpeter equation crucial for the
description of bound states by quantum field theories. Therefore,
it represents a standard tool in hadron physics. It is obtained by
assuming the bound-state constituents to propagate as free
particles (with effective masses) and by considering their
interactions in the `instantaneous limit'.

However, depending on the specific Lorentz behaviour of the
Bethe--Salpeter interaction kernel, implementation of confinement
in the Salpeter equation in too na\"ive ways might lead to the
prediction of unstable states where only truly bound states are
expected: Their energy eigenvalues may be embedded in a continuous
spectrum, for instance.

These observations call for a rigorous spectral analysis of the
Salpeter equation, in order to pin down its essential spectral
features by regarding a bound state as stable if its energy (or
mass) eigenvalue belongs to a real and discrete part of the
spectrum that is bounded from below. Reusing ideas and methods
designed in our previous studies of the `reduced Salpeter
equation' \cite{Lucha07:HORSE,Lucha07:StabOSS-QCD@Work07} and of
its improvement \cite{Lucha05:IBSEWEP(a),Lucha05:IBSEWEP(b),
Lucha05:EQPIBSE} derived by allowing for `dressed' propagators of
all bound-state constituents \cite{Lucha07:SSSECI-Hadron07,
Lucha07:HORSEWEP}, we therefore embark on a thorough investigation
of full --- in contrast to reduced --- Salpeter equations for
fermion--antifermion bound states.

We exploit the serendipity that describing confinement by a
configuration-space potential of harmonic-oscillator form reduces
the Salpeter integral equation to a system of (radial) eigenvalue
differential equations. The analysis of confining interactions of
time-component Lorentz-vector nature is comparatively easy and the
stability of all bound states may be established analytically
\cite{Lucha08-C8} by constructing operator inequalities and
applying the characterization of all discrete eigenvalues of
self-adjoint operators bounded from below by the minimum-maximum
principle \cite{MMP-OI(a),MMP-OI(b),MMP-OI(c),MMP-OI(d),MMP-OI(e),
MMP-OI(f)}. Regrettably, the proof cannot be transferred to the
case of a confining interaction of Lorentz-scalar or -pseudoscalar
structure but for a linear combination of time-component
Lorentz-vector and Lorentz-scalar confining interactions stability
is assured (irrespective of the relative sign of our two
contributions) if the vector dominates the scalar
\cite{Lucha10:QCD@Work10}. `Non-harmonically', we will face
integral equations \cite{Lucha:IntEq}.

\bibliographystyle{aipproc}\end{document}